\documentstyle[epsf,aps,tighten,multicol]{revtex}

\begin{document}

 \title {Multiplicative noise: A mechanism leading to nonextensive 
 statistical mechanics}

 \author{Celia Anteneodo and Constantino Tsallis}

 \address{Centro Brasileiro de Pesquisas Fisicas, 
Dr Xavier Sigaud 150, 22290-180, Rio de Janeiro-RJ, Brazil }

 \maketitle

 \begin{abstract}

 A large variety of microscopic or mesoscopic models lead to generic results 
 that accommodate naturally within Boltzmann-Gibbs statistical mechanics 
 (based on $S_1\equiv -k \int du\, p(u) \ln p(u)$). 
 Similarly, other classes of models point toward nonextensive statistical mechanics 
 (based on $S_q \equiv k [1-\int du\, [p(u)]^q]/[q-1]$, 
 where the value of the entropic index $q\in\Re$ depends on the specific model). 
 We show here a family of models, with multiplicative noise, which belongs to the nonextensive 
 class. More specifically, we consider Langevin equations of the type  
 $\dot{u}=f(u)+g(u)\xi(t)+\eta(t)$, where $\xi(t)$ and $\eta(t)$ are 
 independent zero-mean Gaussian white noises with respective amplitudes $M$ and $A$.  
 This leads to the Fokker-Planck equation 
 $\partial_t P(u,t) = -\partial_u[f(u) P(u,t)] 
 + M\partial_u\{g(u)\partial_u[g(u)P(u,t)]\} + A\partial_{uu}P(u,t)$. 
 Whenever  the deterministic drift is proportional to the noise induced one, i.e., 
 $f(u) =-\tau g(u) g'(u)$,  the stationary solution is 
 shown to be $P(u, \infty) \propto \bigl\{1-(1-q) \beta 
 [g(u)]^2 \bigr\}^{\frac{1}{1-q}}\;$  
 (with $q \equiv \frac{\tau + 3M}{\tau+M}$ and 
 $\beta=\frac{\tau+M}{2A}$). 
 This distribution is precisely the one optimizing $S_q$ with the constraint 
 $\langle [g(u)]^2 \rangle_q \equiv  \{\int du\,[g(u)]^2[P(u)]^q \}/ 
 \{ \int du\,[P(u)]^q \}=\;$constant. 
 We also introduce and discuss various characterizations of the width of the distributions.
 \end{abstract}


 \begin{multicols}{2}

 Ubiquitous systems can be naturally described 
 within Boltzmann-Gibbs (BG) statistical mechanics 
 (based on the entropy $S_1\equiv -k \int du\,p(u) \ln p(u)$). 
 These systems have in common the fact that they like {\it equally} to live everywhere 
 they are allowed to, that is, they are ergodic in the entire phase space.  
 However, there are other systems that, depending on the initial conditions, 
 may prefer a particular subspace. 
 The BG scenario may not be appropriate any longer and an extension of the usual 
 thermostatistical description, taking into account the features of such subspace, 
 would be required. 
 If that subspace has a scale invariant geometry,  
 a hierarchical or multifractal structure, 
 then the model points toward nonextensive statistical mechanics 
 (based on the entropic form 
 $S_q \equiv k\, [1-\int du\, [p(u)]^q]/[q-1]\,,\;q \in \Re$)\cite{Sq} 
 (see \cite{reviews} for reviews). 
 Among the models which belong to this category, one finds low-dimensional 
 dissipative and conservative maps\cite{maps}, 
 fractional and nonlinear Fokker-Planck equations\cite{fFP}, 
 Langevin dynamics with fluctuating temperature\cite{beck}, 
 growth of many-body scale-free networks\cite{networks} and 
 long-range many-body classical Hamiltonians\cite{vito}. 
 The corresponding value of the entropic index $q$ depends on the specific model, 
 or, more precisely, on the nonextensivity universality class of the model. 
 It is our purpose here to show a large family of models with multiplicative noise which
 belongs to the nonextensive class. 
   
 Microscopic dynamics, containing multiplicative noise,  
 may be encountered in many dynamical processes, 
 such as in stochastic resonance \cite{wio1}, 
 noise induced phase transitions \cite{wio2},  
 granular packings \cite{moukarzel},  
 and others \cite{nakao,others}. 
 Due to its significance, stochastic processes with multiplicative noise have been 
 the subject of numerous studies
 in the last decades\cite{multiplicative,vkampen,risken}. 
 Here we will consider processes subject to both  additive and multiplicative 
 noises and described by the dimensionless stochastic differential equation 
 of the form

 \begin{equation}
 \label{langevin_G}
 \dot{u}= f(u) \,+\, g(u)\xi(t) \,+\, \eta(t),
 \end{equation}
 where $u(t)$ is a stochastic variable, $f,g$ are arbitrary functions ($g(0)=0$), 
 and  $\xi(t), \eta(t)$ 
 are  uncorrelated and Gaussian-distributed zero-mean white noises, hence satisfying

 \begin{equation}
 \langle \xi(t)\xi(t') \rangle=2M\delta(t-t'), \;\; 
 \langle \eta(t)\eta(t') \rangle=2A\delta(t-t'), 
 \end{equation}
 where $M,A>0$ are the noise amplitudes and stand for 
 ``multiplicative'' and ``additive'', respectively.  
 Clearly, some degree of correlation between the two noises can be of physical relevance, 
 however, this remains out of the present scope. 
  The deterministic drift $f(u)$ can be interpreted either as a damping force 
 (whenever $u$ is a velocity-like quantity) or as an external force 
 (when motion is 
 overdamped and $u$ represents a position coordinate). Other interpretations 
 are possible as well, depending on the particular system treated. 
 
 It is interesting to note that additive and multiplicative terms in Eq. (\ref{langevin_G}) 
 could be gathered in an effective multiplicative noise term \cite{wio}, 
 however we prefer to keep track of both sources independently.
 
 The stochastic differential equation (\ref{langevin_G}) is not 
 completely defined and must be complemented by an additional rule. 
 This is due to the fact that each pulse of the stochastic noise 
 produces a jump in $u$,  then the question arises: 
 which is the value of $u$ to be used in $g(u)$? 
 This is the well known It\^o-Stratonovich controversy \cite{risken,itosta}. 
 In the It\^o definition, the value before the pulse must be used, whereas 
 in the Stratonovich definition the values before and after the pulse contribute 
 in a symmetric way.  If noise were purely additive, then both definitions agree.

The Fokker-Planck equation for the probability density $P(u,t)$, associated 
to Eq. (\ref{langevin_G}), can be obtained from the Kramers-Moyal expansion 
$\partial_t P =\sum_{n\geq 1} (-\partial_u)^n [D^{(n)}P]$, where the coefficients are
given by
 $D^{(n)}(x,t)=\frac{1}{n!}\lim_{\tau\to 0} \frac{[u(t+\tau)-x]^n}{\tau}|_{u(t)=x}$. 
  These coefficients can be readily obtained following the standard lines found for 
  instance in \cite{risken}. 
 Using the Stratonovich definition of stochastic integral, one gets 
 
  \begin{eqnarray}
 D^{(1)}(u,t) &=& f(u) + M g(u)g'(u) \;\equiv\; J(u), \label{J} \\
 D^{(2)}(u,t) &=& A+M[g(u)]^2        \;\equiv\; D(u),        \label{D}
 \end{eqnarray}
 while $D^{(n)}(u,t)=0$ for $n\geq 3$. Then, one arrives straightforwardly at

 \begin{equation}
 \label{FP_0}
 \partial_t P = -\partial_u j(u) .
 \end{equation}
 where $j(u) \equiv J(u) P - \partial_{u} [D(u) P ]$ is the current. 
 In the It\^o calculation, Eq. (\ref{J}) becomes $J(u)=f(u)$, that is, the 
 noise-induced (or spurious) drift is missing. In what follows we will 
  adopt the Stratonovich definition. However, this choice will not 
 affect the present discussion excepting for a redefinition of some of the 
 parameters that are involved.

 Eq. (\ref{FP_0}) can also be written as

 \begin{equation}
 \partial_t P = -\partial_u( f(u) P) + 
 M\partial_u(g(u)\partial_u[g(u)P]) + A\partial_{uu}P.
 \label{FP_G}
 \end{equation}
 %
 In some processes, the deterministic and noise-induced drifts may have the same functional form. 
 Let us put this condition as follows:

 \begin{equation}
 \label{forcing}
 f(u)=-\tau g(u) g'(u),
 \end{equation}
 $\tau$ being a proportionality constant. 
 In other words, $f(u)$ is derived from a potential-like function $V(u)=\frac{\tau}{2}[g(u)]^2$. 
 Let us note that the particular case $g(u)\propto f(u)\propto u$, which is a natural first 
 choice for a physical system, verifies this condition. However, since no extra calculational 
 difficulties emerge, we will discuss here the more general case (\ref{forcing}). 
 Notice that in the absence of deterministic forcing, condition (\ref{forcing})
 is trivially satisfied for any $g$ by settig $\tau=0$.

In the present paper we will restrict to the stationary solutions for no 
flux boundary conditions (i.e., such that $j(-\infty)=j(\infty)=j(u)=0$), 
although more general conditions could in principle also be considered.
 If  (\ref{forcing}) is verified, then, the stationary solution 
 $P_s(u)$  is of the $q$-exponential form appearing in nonextensive 
 statistical mechanics \cite{Sq}. More precisely, in that case, one obtains

 \begin{equation}
 \label{qexpo_gen}
 P_s(u) \,\propto\, \Bigl[ 1 +  (q-1)\beta  [g(u)]^2  \Bigr]^\frac{1}{1-q},
 \end{equation}
 where $\beta  \equiv \frac{1}{kT} =\frac{\tau+M}{2A}$ (where by $T$ we generically mean 
 the amplitude of an effective noise) and 
 \begin{equation} 
 q=\frac{\tau + 3M}{\tau+M} \;.
 \end{equation}
 As an aside comment let us mention that if we had used the It\^o convention we 
 would have obtained $q=\frac{\tau + 4M}{\tau+2M}$. Let us now go back to our Stratonovich choice. 
 
$P_s$ is normalizable if $|g(u)|$ 
 grows with $|u|$ faster than $|u|^\frac{1}{1+\tau/M}$ ($\tau>-M$).  
 This probability distribution function (pdf) optimizes

 \begin{equation} \label{Sq}
 S_q \equiv k\frac{1-\int du\; [P(u)]^q}{q-1}\, ,
 \end{equation}
 with the constraint 
 \begin{equation}
 \langle [g(u)]^2 \rangle_q \equiv  \frac{\int du\;[g(u)]^2\,[P(u)]^q }{ \int du\;[P(u)]^q }=\;\mbox{ constant}. 
 \end{equation}

 The condition (\ref{forcing}) is not necessary for having solutions of the $q-$exponential form, 
 as it follows along the lines of \cite{lisa,kq97}, in spite of the fact that the models therein 
 considered are different from the present one.

 Let us consider in more detail the case when both the forcing and the 
 multiplicative noise depend on the stochastic variable $u$ as a power law, 
 that is, when the Langevin Eq. (\ref{langevin_G}) becomes

 \begin{equation}
 \label{langevin}
 \dot{u}=-\gamma u|u|^{r-1} \,+\, u|u|^{s-1}\xi(t) \,+\, \eta(t),
 \end{equation}
 with $r,s \ge 0$ and a drift coefficient $\gamma$ typically positive. 
 In this case, the deterministic drift is derived from a confining 
 potential-like function of the form $V(u)=\gamma|u|^{r+1}/(r+1)$. 
 The corresponding Fokker-Planck equation becomes

 \begin{equation}
 \nonumber
 \partial_t P \,=\, -\partial_u(-\gamma u|u|^{r-1} P) \,+\, 
 M\partial_u(|u|^s\partial_u[|u|^sP]) 
\,+\, A\partial_{uu}P.
 \label{FP}
 \end{equation}
 Hence, its stationary solution $P_s(u)$ is given by

 \begin{equation}
 \label{SS}
 P_s(u) \,=\, \frac{ P_o\, {\rm e}^{-h(u)}}{ (1 + \frac{M}{A} |u|^{2s})^{1/2}} \; ,
 \end{equation}
 $P_o$ being the normalization constant and

 \begin{equation}\label{f}
 h(u) \,\equiv\, \frac{\gamma |u|^{1+r}}{A[1+r]} \;_2F_1\Bigl(\frac{1+r}{2s},1;1
 +\frac{1+r}{2s}; -\frac{M}{A} |u|^{2s} \Bigr) ,
 \end{equation}
 where $_2F_1$ is the hypergeometric function.

 We shall analyze now some limiting cases: 
 \\[3mm] 
 (A) For vanishing deterministic forcing ($\gamma\to 0$), 
 Eq. (\ref{SS}) becomes

 \begin{equation}
 P_s(u) \,=\, \frac{ P_o}{ (1 + \frac{M}{A} |u|^{2s})^{1/2}} ,
 \end{equation}
 with
 \begin{equation}
 P_o \,=\, \frac{ (M/A)^{1/(2s)}\;
 \Gamma(\frac{1}{2}) }{ 2\,\Gamma( 1+\frac{1}{2s})\;
 \Gamma(\frac{1}{2}-\frac{1}{2s} ) } \;,
 \;\;\;\;\;\; \mbox{for $s>1$} .
 \end{equation}
 This pdf is of the $q$-exponential form (with $q=3$) as expected because, in this case, 
 condition (\ref{forcing}) is trivially true since it corresponds to $\tau=0$. 
 Interestingly, in the presence of multiplicative noise, the steady state 
 $P_s$ is normalizable {\it even in the absence of a confining potential}, 
 as long as $s>1$. The so called spurious drift which originates from the 
 multiplicative noise term in the Langevin equation is responsible 
 for kicking the system back close to
 the origin, where fluctuations are smaller. 
 On the other hand, the additive noise plays an essential role, 
 providing fluctuations which avoid the full concentration (at 
 the origin) that would occur otherwise.  
 This type of stabilizing effect of the multiplicative noise is long known \cite{stabil}. 
 \\[3mm] 
 (B) In the limit $M \to 0$, all other parameters being fixed, i.e., 
 when no multipliciative noise is present, Eq. (\ref{SS}) becomes of the following stretched exponential form:

 \begin{equation}
 P_s(u)\,=\, \frac{ \bigl[\frac{\gamma}{A[1+r]} \bigr]^\frac{1}{1+r}    }
 { 2\,\Gamma( 1+\frac{1}{1+r}) } 
 \;e^{-\frac{\gamma}{A[1+r]}\, |u|^{1+r} }.
 \end{equation}
 In particular, for the linear forcing ($r=1$), 
 the Gaussian pdf is recovered.
 \\[3mm]  
 (C) In the limit $A\to 0$, i.e., for vanishing strength of the additive noise, 
 the steady state is normalizable for $s<1$ and $p \equiv r+1-2s> 0$. The 
 condition $p>0$ implies that the potential of the drift has to be 
 steep enough to confine the system and yield a steady state. 
 Then, we have

 \begin{equation}
 \label{smallA}
 P_s(u)\,=\, \frac{ p\; \bigl[ \frac{\gamma}{Mp} \bigr] ^\frac{1-s}{p} }
 { 2\;\Gamma( \frac{1-s}{p}) } \;
 \frac{ e^{-\frac{\gamma}{Mp}\,|u|^p}}{|u|^s} \;.
 \end{equation}
 Vanishing $A$ concentrates the probability at the origin. 
 Again, the Gaussian distribution is recovered 
 for $r=1$ (harmonic forcing) and $s=0$ (additive noise). \\

 The limits analyzed above are not generically interchangeable, 
 that is, convergence is not necessarily uniform.

 In the general case where $(A,M,\gamma)$ are all finite, the same dependence on $u$ as 
 that in Eq. (\ref{smallA}) is obtained for sufficiently large $|u|$. 
 In fact, if one defines the dimensionless variable 
 $\bar{u}=u/\lambda$ with $\lambda \equiv (A/M)^\frac{1}{2s}$, it is clear that the 
 asymptotic expression of Eq. (\ref{SS}) for $|\bar{u}|\to \infty$ corresponds 
 to both limits $|u|\to \infty$ and $A\to 0$.

 For finite $(A,M,\gamma)$, in the $p=0$ marginal case where the  drift 
 and multiplicative-noise exponents are related through $1+r=2s$ 
 (hence condition (\ref{forcing}) is verified with $\tau=\gamma/s$),  a $q$-exponential pdf
 emerges at asymptotically long time. 
 In fact, in that case,  the hypergeometric in Eq. (\ref{f}) becomes
 $_2F_1(1,1;2;-z)=\ln(1+z)/z$ and one gets

 \begin{equation}
 \label{qexpo}
 P_s(u) \,=\, \frac{ P_o}{ (1 +  |u/\lambda|^{2s} )^\frac{1}{q-1} }\;,
 \end{equation}
 with
 \begin{equation} 
 q= \frac{(\gamma/s)+3M}{(\gamma/s)+M} 
 \end{equation}
 and
 \begin{equation}
 P_o \,=\, \frac{ \Gamma(\frac{1}{2} +\frac{\gamma}{2sM}) }
 { 2\lambda \,\Gamma( 1+\frac{1}{2s})\;
 \Gamma(\frac{1}{2} +\frac{\gamma}{2sM} -\frac{1}{2s} ) } \;\;.
 \end{equation}
 $P_s(u)$  is normalizable for $s+\gamma/M>1$. Notice that
 $q$-exponential pdf's can appear even for negative $\gamma$ 
 (i.e., for repulsive deterministic forces). 
 The $1+r=2s$ class includes the particular linear case $r=s=1$, that has already been 
 treated in the literature \cite{nakao,sakaguchi}. 
 In Fig.~1 we exhibit the steady state pdf (\ref{qexpo}) for several values 
 of the system parameters.

 Assuming that $\langle u\rangle =0$, the width of $P(u,t)$ can be 
 characterized in many ways, such as 
 (i) the inverse of the height at the origin, namely $1/P(0,t)$; 
 (ii) the width at half height, $\Delta$; 
 (iii) the square root of the mean value of $u^2$; 
 (iv) the $2s$-root of the mean value of $|u|^{2s}$;  
 (v) the square root of the $q$-expectation values of $u^2$;
 (vi) the $2s$-root of the $q$-expectation value of $|u|^{2s}$, or even 
 combinations of these, for instance (vii) $\sqrt{\Delta/P(0,t)}$ (see also \cite{prato}). 
 Let us recall that the normalized $q$-expectation of $\psi$ is defined as

 \begin{equation}
 \label{qexpectation}
 \langle \psi \rangle_q(t) \equiv  \frac{\int du\;\psi(u) [P(u,t)]^q } 
 { \int du\;[P(u,t)]^q }  \;,
 \end{equation}
 hence, $\langle \psi \rangle_1$ 
 equals the usual mean value $\langle \psi \rangle$.
  
 For the usual case where $P(u,t)$ is a Gaussian, all these definitions  
 basically coincide. This is not so in general, as we shall illustrate 
 in what follows for the stationary state of Eq. (\ref{qexpo}).

 (i) The inverse of the height at the origin:
 \begin{equation}
 \frac{1}{P_s(0)} \,=\, \lambda \;
 \frac{ 2\,\Gamma( 1+\frac{1}{2s})\;\Gamma(\frac{1}{2} 
 +\frac{\gamma}{2sM} -\frac{1}{2s} ) }
 { \Gamma(\frac{1}{2} +\frac{\gamma}{2sM}) } .
 \end{equation}

 (ii) The width at half height:
 \begin{equation}
 \Delta \,=\, 2\lambda  \,(2^{1/(\frac{\gamma}{2sM}
 +\frac{1}{2})} -1)^\frac{1}{2s}.
 \end{equation}

 (iii) The square root of the mean value of $u^2$:
 \begin{equation}
 \langle  u^2\rangle^{1/2} \,=\, \lambda
 \;\sqrt\frac{ \Gamma( \frac{3}{2s})  \;
 \Gamma(\frac{1}{2} +\frac{\gamma}{2sM} -\frac{3}{2s} ) }
 { \Gamma( \frac{1}{2s}) \;\Gamma(\frac{1}{2} +\frac{\gamma}{2sM} -\frac{1}{2s} ) },
 \end{equation}
 for $\gamma/M>3-s$ (otherwise it diverges).

 (iv) The $2s$-root of the mean value of $|u|^{2s}$,  
 \begin{equation}
 \langle |u|^{2s}\rangle^{1/2s} \,=\,  
 \; \frac{\lambda}{  \bigl(  \gamma/M-s-1 \bigr)^\frac{1}{2s} }  
 \end{equation}
 for $\gamma/M>1+s$ (otherwise it diverges).

 (v) The square root of the $q$-expectation value of $u^2$:  
 \begin{equation}
 \langle u^2\rangle_q^{1/2} \,=\, \lambda
 \;\sqrt\frac{ \Gamma( \frac{3}{2s})  \;\Gamma(\frac{3}{2} +\frac{\gamma}{2sM} -\frac{3}{2s} ) }
      { \Gamma( \frac{1}{2s}) \;\Gamma(\frac{3}{2} +\frac{\gamma}{2sM} -\frac{1}{2s} ) },
 \end{equation}
 for $\gamma/M>3(1-s)$ (otherwise it diverges), 
 and

 (vi) The $2s$-root of the $q$-expectation value of $|u|^{2s}$
 \begin{equation}
 \langle |u|^{2s}\rangle_q ^{1/2s} \,=\, 
 \; \frac{\lambda}{  \bigl(  \gamma/M + s-1 \bigr)^\frac{1}{2s} }  ,
 \end{equation}
 for $\gamma/M>1-s$. Notice that this condition has already been encountered. Indeed, 
 it is necessary for normalizability. 
 An important remark is mandatory: of all the above characterizations of width which are 
 based on expectation values (i.e., $\langle u^2 \rangle^{1/2}$, 
 $\langle |u|^{2s}\rangle^{1/2s}$, $\langle u^{2}\rangle_q^{1/2}$  and 
 $\langle |u|^{2s}\rangle_q^{1/2s}$ ), {\it only the last one does not diverge 
 in any admissible (i.e., normalizable) case}. This is particularly remarkable because this 
 generalized expectation value is, as already seen, precisely the one to be used as constraint 
 in the optimization of $S_q$. In addition to this, it is worthy noticing that 
 $\langle |u|^{2s}\rangle_q^{1/2s}$ is in all cases comparable to $\sqrt{\Delta/P_s(0)}$, 
 a semi-empirical quantity which takes into account the contributions  of both body  and tails.

 In Fig.~2 we exhibit all these quantities as a function of  
 $M/\gamma$ for typical values of $s$ with $r=2s-1$. 
 As $M/\gamma$ increases, in all cases, $P_s(u)$ becomes more peaky around 
 the origin ($\Delta$ decreases).  
 Moreover, for increasing $M/\gamma$ and $s=1$,  
 the tail tends to $1/|u|$ (hence $P_s(0)\to 0$). Also for $s=1/2$, $P_s(0)\to 0$ 
 as the pdf becomes more tailed, but $M/\gamma$ has the upper bound 2. 
 Contrarily to the previous cases, for $s>1$, as $M/\gamma$ increases, 
 $P_s(0)$ also increases.

 The $q$-exponential character is not exclusive of the steady state but it also 
 emerges along the time evolution of the pdf, as illustrated in Fig. 3. 
 In this figure we employ the semi-$\ln_q$ representation, where 
 $\ln_q x= (x^{1-q}-1)/(q-1)$.  
 Notice in Figs. 3.b and 3.c that the curve for $t\to \infty$ is a straight line. 
 In 3.c, curves are almost straight lines, thus indicating that the pdf's are very close to  
 $q$-exponentials along time.

 In the presence of multiplicative noise, the system variables directly couple  to 
 noise. Therefore, behaviors are observed that can not occur 
 in the presence of additive noise alone. 
 In particular, $q$-exponential pdf's of the form given by Eq. (\ref{qexpo_gen}) can  arise. 
 For the present processes, $q$-exponentials do occur either
 in the absence of forcing ($\gamma=0)$ or   
 for drifts verifying Eq. (\ref{forcing}), which, for the power-law case,
 becomes  $r=2s-1$.
 The deterministic forcing does not need to be confining (i.e., 
 $\gamma<0$ is allowed) for the formation of a $q$-exponential stationary pdf, 
 provided $\gamma/M>1-s$.
 In any case, if a repulsive effective force prevails for some period of time, the 
 variable $u$ can take large values and power law tails can arise. 
 In particular, a $q$-exponential occurs when the stochastic forcing is proportional to the deterministic one. 
 Alternatively, during the intervals when the effective force is attractive, 
 the probability tends to concentrate at the origin. Then, at this stage, the additive noise plays 
 a fundamental role allowing the existence of a normalizable steady state 
 by avoiding collapse of the pdf at the origin. 
 When both kinds of noise occur simultaneously, the presence of multiplicative 
 noise can not be formally avoided by a simple transformation of variables. 
 The particular interplay between additive and  
 multiplicative noises as well as that between deterministic and stochastic drifts 
 can lead to the appearance of $q$-exponentials.

 It is worthy to emphasize at this point that the stationary solutions of the present 
 problem have the general form given by Eq. (\ref{SS}). The $q$-exponential pdf's 
 represent a special case, 
 which in turn includes the Boltzmann-Gibss pdf as an even more special one, 
 corresponding to the standard thermal equilibrium.
 Furthermore, the $q$-exponential pdf's are unique in the sense that they optimize, 
 under appropriate constraints, the entropy functional $S_q$ (Eq. (\ref{Sq})). 
 This entropic form is in turn unique  in the sense that it is the only one which satisfies\cite{axioms} 
 a set of conditions naturally generalizing both the Shannon and the Kinchin axioms. 
 Also, $S_q$ is consistent with stability (or robustness) of the $q$-exponential pdf's in the sense 
 described by Abe\cite{stability}, whereas Renyi entropy and the normalized version of $S_q$ are not.

 Let us conclude by stressing that 
 $q$-exponentials have also been observed in a variety of
 similar processes \cite{beck,lisa,annunziato}. 
 The mechanism leading to such distributions is expected to be present in 
 systems with long--range memory, long-range interactions, fractal or hierarchical structures 
 and similar scenarios.

 In what concerns the model defined by Eqs. (3)-(5), it is clear that $f(u)$ and $g(u)$ are 
 generically independent functions. However, it does occur that when they are connected through Eq. 
 (\ref{forcing}), the $q$-exponential form emerges naturally. It would no doubt be very 
 interesting if we had a geometrical interpretation of this fact. Hints along this line 
 would be welcome.

 As illustrated in Fig. 3 (deterministic drift proportional to the stochastic one), 
 the $q$-exponential is exact for $t\to \infty$ and it is an excellent approximation $\forall t$.

 The best characterizations of the width of the pdf's clearly 
 are those which remain finite under generic circumstances. In our case, 
 this happens for $1/P(0,t)$, $\Delta$, 
 $\langle |u|^{2s}\rangle_q ^{1/2s}$ or combinations such as 
 $\sqrt{\Delta/P(0,t)}$. The latter two are preferable since they contain attributes of both body and 
 tails of the pdf. For a mathematically convenient and generic characterization, 
 clearly $\langle |u|^{2s}\rangle_q ^{1/2s}$ is the most appropriate.

 We are grateful to C.F. Moukarzel for useful remarks 
 and to H. Wio with regard to the comment in \cite{wio}. 
 We also acknowledge PRONEX, FAPERJ and CNPq (Brazilian agencies) for partial financial support.

 \vspace*{1cm}

 \centerline{{\bf FIGURE CAPTIONS}}

 \bigskip
 \noindent
 Fig. 1.~  Steady state pdfs for s=0.5 (a), 1 (b), and 2 (c), 
 with $r=2s-1$, $A/\gamma=1$ and 
 different values of $M/\gamma$ indicated in the figure. 
 Notice that the alternative representation $\lambda \,P_s(u)$ vs. $u/\lambda$ 
 with $\lambda=(A/M)^{1/2s}$ would give profiles independent on the particular 
 choice of $A/\gamma$.
 \vfill

 \bigskip
 \noindent
 Fig. 2.~  Different parameters characterizing the ``width'' of the distribution  
 as a function of $M/\gamma$ for s=0.5 (a), 1 (b), and 2 (c),  with $r=2s-1$ 
 and $A/\gamma=1$. For $\gamma >0$, normalizabity implies $M/\gamma < 1/(1-s)$ 
 if $s < 1$; for $s\ge 1$, the physical region of $M/\gamma$ extends up to infinity.
 \vfill

 \bigskip
 \noindent
 Fig. 3.~  Time evolution of the pdf for $r=s=1$, $A=\gamma=1$ and $M=0.2$
  at different times indicated in the figure, in different representations: 
 linear (a), semi-$\ln_q$  
 with $q=\frac{\gamma+3M}{\gamma+M}$ (b), and semi-$\ln_{q'}$, with time 
 dependent $q'$ (c). 
 We used the ansatz $P=P_o (1+(q'-1)\beta u^2)^{1/(1-q')}$, with 
 $P_o=\sqrt{\frac{\pi}{\beta(q'-1)}}
 \frac{\Gamma(\frac{1}{q'-1}-\frac{1}{2})}
 {\Gamma(\frac{1}{q'-1})}$.
 In the inset we present $q'$ and $\beta$ 
 vs. $t$. In  particular, $q(0)=1$, $1/\beta(0)=0$, 
 $q(\infty)= 4/3$ and $1/\beta(\infty)=5/3$.
 \vfill
 
 \end{multicols}

\newpage
\begin{figure}[htb] 
\unitlength 1mm 
\vspace*{-2cm}
\begin{center} 
 {\epsfxsize 11.0cm \epsfbox{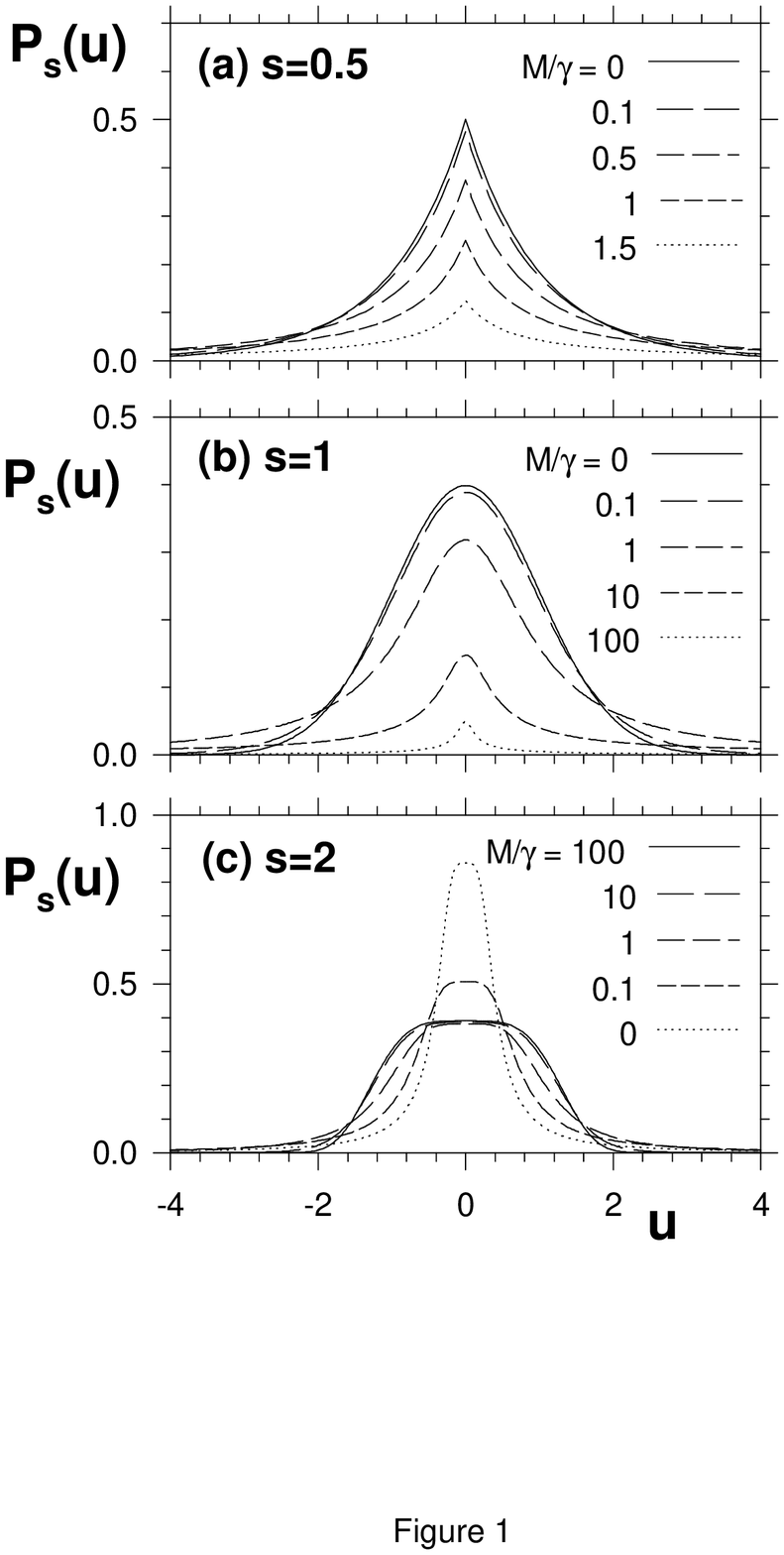}} 
\end{center} 
\vspace*{-6cm}
\end{figure}

\newpage 
 \begin{figure}[htb] 
\unitlength 1mm 
\vspace*{-1cm}
\begin{center} 
 {\epsfxsize 8.5cm \epsfbox{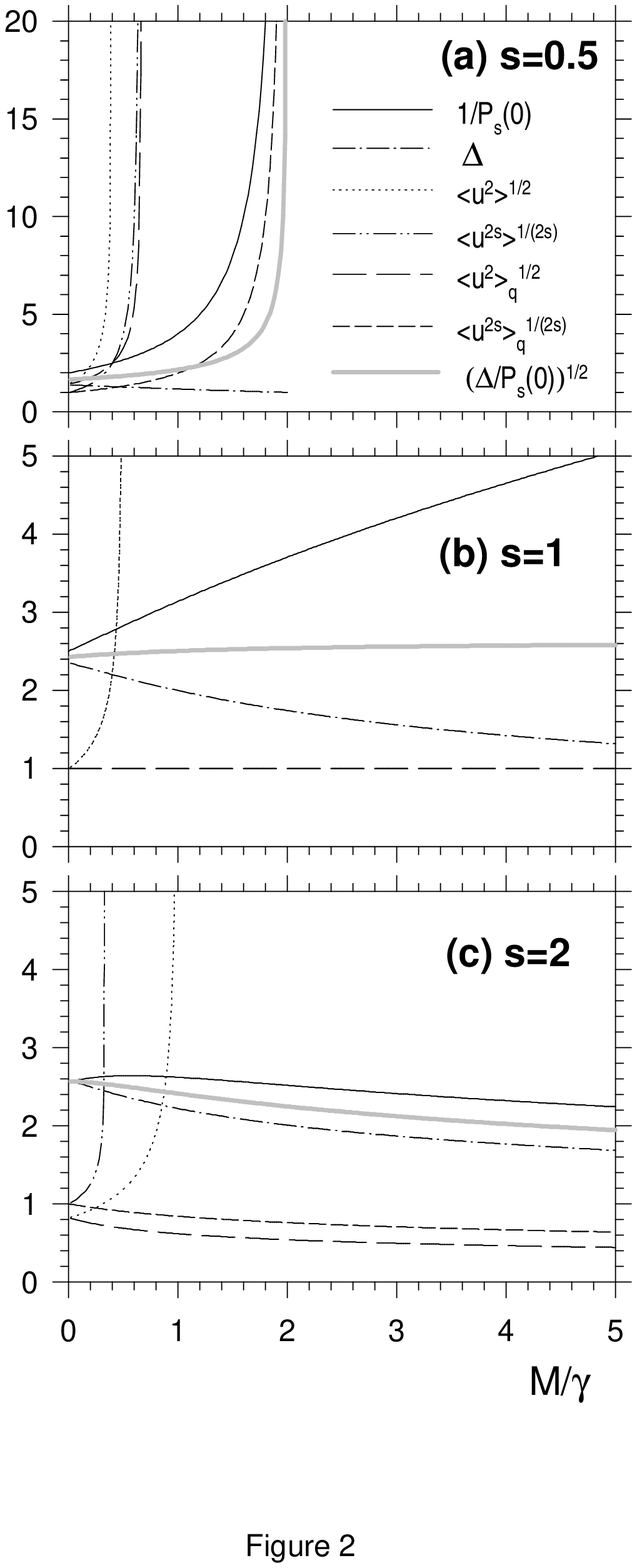}} 
\end{center} 
\vspace*{-4cm}
\end{figure}

\newpage
\begin{figure}[htb] 
\unitlength 1mm 
\vspace*{-0cm}
\begin{center} 
 {\epsfxsize 10.0cm \epsfbox{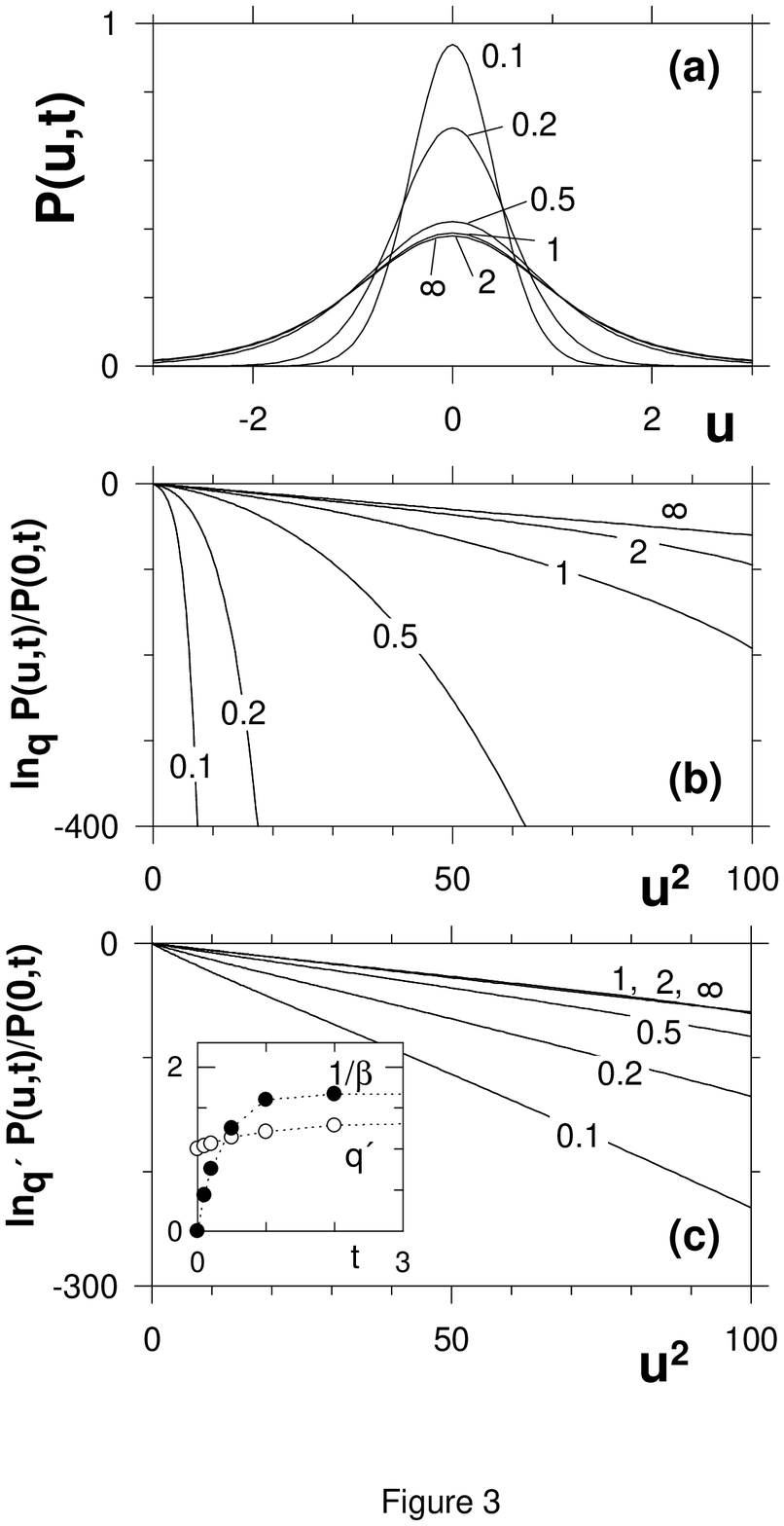}} 
\end{center} 
\vspace*{-2cm}
\end{figure}


\begin{thebibliography}{99}



 \bibitem{Sq}
 C. Tsallis, J. Stat. Phys. {\bf 52}, 479 (1988);
 E.M.F. Curado and C. Tsallis, J. Phys. A {\bf 24}, L69 (1991)
 [Corrigenda: {\bf 24}, 3187 (1991) and {\bf 25}, 1019 (1992)];
 C. Tsallis, R.S. Mendes and A.R. Plastino, Physica A {\bf 261}, 534 (1998).


 \bibitem{reviews}S.R.A. Salinas and C. Tsallis, eds., {\it Nonextensive Statistical Mechanics 
 and Thermodynamics}, Braz. J. Phys. 
 {\bf 29} (1999); S. Abe and Y. Okamoto, eds.,
 {\it Nonextensive Statistical Mechanics and Its Applications}, Series {\it Lecture Notes in Physics} 
 (Springer-Verlag, Berlin, 2001); G. Kaniadakis, M. Lissia and A. 
 Rapisarda, eds., {\it Non Extensive Thermodynamics and Physical 
 Applications},  Physica A {\bf 305} (Elsevier, Amsterdam, 2002); 
 M. Gell Mann and C. Tsallis, eds., {\em Nonextensive Entropy--Interdisciplinary 
 Applications}, (Oxford University 
 Press, Oxford, 2002), in press.
 A regularly updated bibliography on the subject is accessible at 
 http://tsallis.cat.cbpf.br/biblio.htm 


 \bibitem{maps} 
 C. Tsallis, A.R. Plastino and W.M. Zheng, Chaos, Solitons and 
                Fractals {\bf 8}, 885 (1997); M.L. Lyra and C. Tsallis, Phys. Rev. Lett. {\bf 80}, 53 (1998); 
 U. Tirnakli, Phys. Rev. E {\bf 62}, 7857 (2000); 
 E.P. Borges, C. Tsallis, G.F.J. Ananos and P.M.C. de Oliveira, Phys. Rev. Lett. {\bf 89}, 254103 (2002); 
 Y.S. Weinstein, S. Lloyd and C. Tsallis, Phys. Rev. Lett. {\bf 89}, 214101 (2002). 



 \bibitem{fFP}  A.R. Plastino and A. Plastino, Physica A  {\bf 222}, 347 (1995); 
                L.C. Malacarne, R.S. Mendes, I.T. Pedron and E.K. Lenzi, Phys. Rev. E {\bf 63}, 030101 (2001); 
                E.K. Lenzi, C. Anteneodo and L. Borland,  Phys. Rev. E {\bf 63}, 051109 (2001);
                I.T. Pedron, R.S. Mendes, L.C. Malacarne and E.K. Lenzi, Phys. Rev. E {\bf 65}, 041108 (2002); 
                L.C. Malacarne, R.S. Mendes, I.T. Pedron and E.K. Lenzi, Phys. Rev. E {\bf  65}, 052101 (2002).


 \bibitem{beck}G. Wilk and Z. Wlodarczyk, Phys. Rev. Lett. {\bf 84}, 2770 (2000); 
 C. Beck, Phys. Rev. Lett. {\bf 87}, 180601 (2001); 
 C. Beck and E.G.D. Cohen, Physica A {\bf 321}, 267 (2003). 


 \bibitem{networks} R. Albert and A.L. Barabasi, Phys. Rev. Lett. {\bf 85}, 5234 (2000).


 \bibitem{vito}  V. Latora, A. Rapisarda and C. Tsallis, Phys. Rev. E {\bf  64}, 056134 (2001).
 
 \bibitem{wio1} C.J. Tessone and H.S. Wio, Mod. Phys. Lett B {\bf 12}, 1195 (1998).


 \bibitem{wio2} S.E. Mangioni, R.R. Deza, R. Toral and H. Wio, Phys. Rev. E {\bf 61}, 223 (2000).


 \bibitem{moukarzel}  C.F. Moukarzel, J. Phys.: Condens. Matter {\bf 14}, 2379 (2002).


 \bibitem{nakao} H. Nakao, Phys. Rev. E {\bf 58}, 1591 (1998).


 \bibitem{others} J.M. Deutsch, Physica A {\bf 208}, 445 (1994); 
                  Y.-C. Zhang, Phys. Rev. Lett. {\bf 56}, 2113 (1986).


\bibitem{multiplicative} R.F. Fox, J. Math. Phys. {\bf 13}, 1196 (1972);
                         N.G. Van Kampen, Phys. Rep. {\bf 24}, 171 (1976);
                         R. F. Fox, Phys. Rep. 48, 179 (1978);
                         L. Arnold, W. Horsthemke and R. Lefever, Z. Phys. B {\bf 29}, 367 (1978).
                         A. Schenzle and H. Brand, Phys. Rev. A {\bf 20}, 1628 (1979).
                         
 \bibitem{vkampen}  N. G. van Kampen, Phys. Rep. {\bf 24}, 171 (1976);
{\it Stochastic Processes in Physics and Chemistry} (North-Holland, Amsterdam, 1981).
                        
 \bibitem{risken} H. Risken, {\em The Fokker-Planck equation. Methods of solution 
 and applications} (Springer-Verlag, New York, 1984). 


\bibitem{wio}    Eq. (\ref{langevin_G}) can be rewritten as 
$ \dot{u}= f(u) + \tilde{g}(u)\zeta(t) $,  where $\tilde{g}(u)=\sqrt{(A+M[g(u)]^2)/C} $  
and $\zeta(t)$ is a gaussian white noise 
satisfying $ \langle \zeta(t)\zeta(t') \rangle=2C\delta(t-t')$, with $C>0$. 



\bibitem{itosta} N.G. Van Kampen, J. Stat. Phys. {\bf 24}, 175  (1981). 


 \bibitem{lisa} L. Borland, Phys. Lett. A {\bf 245}, 67 (1998);
                T. Munakata and S. Mitsuoka, J. Phys. Soc. Jap. {\bf 69}, 92 (2000).


 \bibitem{kq97} G. Kaniadakis and P. Quarati, Physica A {\bf 237}, 229 (1997). 


 \bibitem{stabil} R. Graham and A. Schenzle, Phys. Rev. A {\bf 26}, 1676 (1982).


 \bibitem{sakaguchi}  H. Sakaguchi, J. Phys. Soc. Japan {\bf 70}, 3247 (2001).


 \bibitem{prato}  C. Budde, D. Prato and M. Re, Phys. Lett. A {\bf 283}, 309 (2001).


 \bibitem{axioms}   R.J.V. Santos. J. Math. Phys. {\bf 38}, 4104 (1997); 
                    S. Abe, Phys. Lett. A {\bf 271}, 74 (2000).


 \bibitem{stability} S. Abe, Phys. Rev. E {\bf 66}, 046134 (2002).


 \bibitem{annunziato} M. Annunziato, Phys. Rev. E {\bf 65}, 021113 (2002).


                                                                  


 \end{thebibliography}
 \end{document}